\documentclass[journal]{IEEEtran}
\ifCLASSINFOpdf
\else
\fi
\usepackage{url}
\usepackage{amsmath,amssymb,amsfonts}
\usepackage{algorithm,algorithmic}
\usepackage{graphicx}
\usepackage{makecell}
\usepackage{textcomp}
\usepackage{xcolor}

\begin{document}
%
\title{Optimal Sizing of Stand-alone Solar PV Systems via Automated Formal Synthesis}
%
%
%

\author{Alessandro~Trindade and Lucas~Cordeiro
\thanks{A. Trindade is with the Department of Electricity, Federal University of Amazonas, Manaus, Brazil, e-mail: alessandrotrindade@ufam.edu.br.}
\thanks{L. Cordeiro is with School of Computer Science, The University of Manchester, UK, e-mail: lucas.cordeiro@manchester.ac.uk.}
\thanks{The work of A. Trindade was supported in part by Newton Fund under Grant $261881580$, and in part by FAPEAM- Amazonas Foundation for Research under Grant PROTI-Pesquisa $2018$.}
}

%
%

\markboth{arXiv, September~2019}%
{Trindade and Cordeiro: Optimal Sizing of Stand-alone Solar PV Systems via Automated Formal Synthesis}
%



\maketitle

\begin{abstract}
There exist various methods and tools to size solar photovoltaic systems; however, these tools rely on simulations, which do not cover all aspects of the design space during the search for optimal solution. In prior studies in optimal sizing, the focus was always on criteria or objectives. Here, we present a new sound and automated approach to obtain optimal sizing  using an unprecedented program synthesis. Our variant of counterexample guided inductive synthesis (CEGIS) approach has two phases linking the technical and cost analysis: first we synthesize a feasible candidate based on power reliability, but that may not achieve the lowest cost; second, the candidate is then verified iteratively with a lower bound cost via symbolic model checking. If the verification step does not fail, the lower bound is adjusted; and if it fails, a counterexample provides the optimal solution. Experimental results using seven case studies and commercial equipment data show that our synthesis method can produce within an acceptable run-time the optimal system sizing. We also present a comparative with a specialized simulation tool over real photovoltaic systems to show the effectiveness of our approach, which can provide a more detailed and accurate solution than that simulation tool.
\end{abstract}


%
\IEEEpeerreviewmaketitle

\section{Introduction}

\IEEEPARstart{L}{ack} of access to clean and affordable energy is considered a core dimension of poverty~\cite{Hussein2012}. Progress has been made worldwide; in particular, the number of people without electricity access fell below $1$ billion threshold for the first time in $2017$~\cite{IEAweo2018}. Decentralized systems led by solar photovoltaic (PV) in off-grid and mini-grid systems are the lowest-cost solution for three-quarters of the additional connections needed to provide universal electricity~\cite{Hussein2012}; specifically, grid extension are the standard in urban areas~\cite{IEAweo2018}.

%
To simulate or evaluate a PV system, there exist various specialized tools, e.g., RETScreen, and HOMER~\cite{Pradhan,Swarnkar}; and even general-purpose simulation tools, as  MATLAB~\cite{Benatiallah2017}. However, these tools rely on simulation, with the drawback of an incomplete coverage since verification of all possible combinations and potential failures of a system is unfeasible~\cite{ClarkeHV18}. 

Optimization of PV systems is not a recent topic; since the $90$'s different techniques using a wide variety of criteria to find ultimate combinations for design parameters, based on intuitive, numerical, and analytical methods were developed and evaluated~\cite{Applasamy2011}. An ideal combination of any PV system consists of the best compromise between two objectives, which is power \textit{reliability} and \textit{system cost}~\cite{Alsadi2018}.

Formal methods could offer great potential to obtain a more effective design process for PV systems~\cite{ClarkeHV18}. In $2012$, a Chinese smart grid implementation considered a case study to address the verification problem for performance and energy consumption~\cite{Yukseletall2012}. In $2015$, an automated simulation-based verification technique was applied to verify the correctness of power system protection settings~\cite{Sengupta2015}. In $2017$, a researcher suggested the application of formal methods to verify and control the behavior of computational devices in a smart grid ~\cite{Abate2017}. Finally, in $2018$, a verification methodology was applied to PV panels and its distributed power point tracking~\cite{Driouich2018}. However, prior studies did not deal with electricity generation or even solar PV systems optimization. Formal methods based on \textit{symbolic model checking} and its application to synthesize PV systems are still unexplored in literature.
 
Here, we have developed a variant of counterexample guided inductive synthesis (CEGIS) for synthesizing optimal sizing of stand-alone PV systems using commercial equipment data. Given a correctness specification $\sigma$, our method uses that as a starting point and then iteratively produces a sequence of candidate solutions that satisfy $\sigma$, related to power reliability. In particular, in each iteration, we synthesize the sizing of stand-alone PV systems, but that may not achieve the lowest cost. The candidate solution is then verified via symbolic model checking with a lower bound that serves as the minimum cost of reference; if the verification step does not fail, the lower bound is adjusted. If it fails, then a counterexample is provided with an optimal sizing that meets both power reliability and system cost. Note that in this study, our focus is not on new criteria or even optimization objectives. Instead, our novelty relies on an effective approach to the pursuit of the optimal solution of PV systems using formal methods. 

In summary, this paper makes the following original contributions: (i) It is the first application of a sound and automated formal synthesis approach, which can provide accurate results of optimal sizing of stand-alone PV systems; (ii) We propose a variant CEGIS method with striking differences of how the {\sc Synthesize} and {\sc Verify} phases work together, with the abolition of the feasible solution candidate vector and the use of an incremental, iterative loop to reach the optimal cost solution of the system; and (iii)  Experimental results with seven case studies show that formal synthesis approach qualitatively outperforms an existing state-of-the-art simulation tool. Our solution is far detailed and closer to the commercial reality (real PV systems) than the solution presented by simulation.



\section{Sizing and Optimization of PV Systems}
\label{sec:Background}

\begin{figure}[h]
\includegraphics[width=0.4\textwidth]{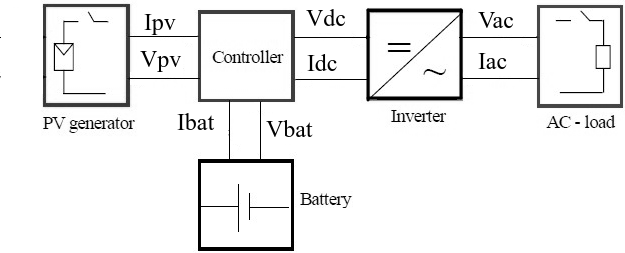}
\centering
\caption{Block diagram for a typical stand-alone PV system~\cite{Hansen}.}
\label{fig:blockdiagram} 
\end{figure}

Fig.\ref{fig:blockdiagram} illustrates a stand-alone PV system block diagram. The PV generator (a panel or an array) is a semiconductor device that can convert solar energy into DC electricity. For night hours or rainy days, we hold batteries where power can be stored and used. The use of batteries as a storage form implies the presence of a charge controller~\cite{Hansen}. The PV arrays produce DC, and therefore when the PV system contains an AC load, a DC/AC conversion (inverter) is required. The AC load dictates the AC electrical load behavior from the house.

In this section, we explain the adopted method to size stand-alone solar PV systems, and what criteria and technique are possible to apply for the optimal sizing. 

\subsection{Sizing Stand-alone Solar PV Systems}
\label{sec:sizing}

We adopted a critical period solar energy method~\cite{Pinho}; in particular, maximum power point tracking (MPPT) charge controller, which is the most common nowadays. First, we need to correct the energy consumption estimated to the load ($E_{consumption}$), which is carried out by Eq.~\eqref{eq:Ecorrected}, where the efficiency of batteries ($\eta_{b}$), controller ($\eta_{c}$), and inverter ($\eta_{i}$) are considered~\cite{Pinho} as follows

\begin{equation}
\label{eq:Ecorrected}
E_{corrected} = \dfrac{E_{consumption}}{\eta_{b} \eta_{c} \eta_{i} }.
\end{equation}

We also need to estimate the energy that each panel produces, called $E_{p}$, in Wh, defined as

\begin{equation}
\label{eq:Ep}
E_{p} = Solar\_Irradiance \times Panel\_Area \times \eta_{p} \times 1000,
\end{equation}

\noindent where the solar irradiance is expressed in terms of $kWh/m^{2}$ and depends on the site where we deploy the PV system; 
the PV panel area is given in $m^{2}$ and corresponds to the size of one PV panel, and $\eta_{p}$ represents the PV panel efficiency.
We compute the total minimum number of needed solar panels ($N_{TPmin}$) as

\begin{equation}
\label{eq:NTPmin}
N_{TPmin} = \dfrac{E_{corrected}}{E_{p}}.
\end{equation}

Notably, we give the total number of panels in series ($N_{PSmin}$) and parallel ($N_{PPmin}$) by

\begin{equation}
\label{eq:NPSmin}
\dfrac{V_{mppt,min}}{V_{maxPower,TempMax}} \leq N_{PSmin} \leq \dfrac{V_{mppt,max}}{V_{maxPower,TempMin}},
\end{equation}

\begin{equation}
\label{eq:NPPmin}
N_{PPmin} = \dfrac{P_{total}}{Number\,Panels\,Series \times P_{max,ref}},
\end{equation}

\noindent where $V_{mppt,max}$ is the maximum operation voltage and $V_{mppt,min}$ 
is the minimum operation voltage of the charge controller; $V_{maxPower,TempMax}$ and 
$V_{maxPower,TempMin}$ are the maximum power voltage from the PV module considering 
the maximum and minimum operational temperature, respectively; 
$P_{total}$ is the total power demanded from the PV system and 
$P_{max,ref}$ is the power supplied from one PV panel in $Watts$.
Regarding batteries, we must first define the total capacity of the battery bank as
\begin{equation}
\label{eq:Cbank}
C_{bank} = \dfrac{E_{corrected} \times autonomy}{V_{system} \times DOD},
\end{equation}

\noindent where the variable $autonomy$ is a design definition and typically has a value ranging from $6$ to $48$h; $ V_{system} $ 
is the DC voltage of the bus, and $ DOD $ is the battery deep of discharge (considered of maximum of 25\% here).
Second, we compute the total (minimum) number of batteries as 
\begin{equation}
\label{eq:Nbtotal}
N_{B}total = N_{BS}min \times N_{BP}min
\end{equation}

\begin{equation}
\label{eq:Nbtotal2}
N_{B}total = \dfrac{V_{system}}{V_{bat}} \times \dfrac{C_{bank}}{1 \,Battery \, Capacity}.
\end{equation}

Regarding the charge controller, it must initially meet the voltage requirement of the PV system, as described by Eq.~\eqref{eq:vcvsystem} to the charge controller voltage: 

\begin{equation}
\label{eq:vcvsystem}
V_{c} = V_{system}.
\end{equation}

The short circuit reference information from the manufacturer's solar panel must be corrected 
to the cell temperature because the field temperature is higher than the nominal or laboratory temperature, and PV system is temperature dependent, as 

\begin{equation}
\label{eq:iscamb}
I_{sc,amb} = \dfrac{G}{G_{ref}} \left[ I_{sc,ref} + \mu_{I} \times (T-25) \right]. 
\end{equation}

The controller must meet the maximum current from the PV array given by Eqs.~\eqref{eq:icmin} and~\eqref{eq:icicmin} as

\begin{equation}
\label{eq:icmin}
I_{c,min} = I_{sc,amb} \times N_{PP},
\end{equation}

\begin{equation}
\label{eq:icicmin}
I_{c} \geq I_{c,min}.
\end{equation}

The inverter sizing check is performed using three equations. Eq.~\eqref{eq:vindc} ensures that 
the input voltage of the controller meets the system voltage. Eq.~\eqref{eq:voutac} ensures that the 
output voltage of the controller meets the AC voltage of the load. Finally, Eq.~\eqref{eq:invcheck} ensures that 
the controller can support the total demand of the load ($Demand$) and the surge power ($P_{surge}$), 
where $V_{in}DC$ is the nominal input voltage and $V_{out}AC$ is the nominal output voltage of the inverter; 
$MAX_{AC,ref}$ is the peak power that the inverter can support.

\begin{equation}
\label{eq:vindc} 
V_{in}DC = V_{system}.
\end{equation}

\begin{equation}
\label{eq:voutac} 
V_{out}AC = V_{AC}.
\end{equation}

\begin{equation}
\label{eq:invcheck} 
\left[ (Demand \leq P_{AC,ref}) \, and \, (P_{surge} \leq MAX_{AC,ref}) \right].
\end{equation}

\subsection{PV Systems Optimization: Criteria and Techniques}

We need to evaluate \textit{power reliability} and \textit{system cost analysis} for the underlying system to select an optimal PV system to meet sizing constraints. An ideal combination of any PV system consists of the best compromise between these two objectives or criteria.

During the PV system design, one of the most important aspects to ensure power system reliability is to analyze power supply availability~\cite{Alsadi2018}. The reason is that solar energy production is intermittent and, therefore, the energy generated usually does not match with the load demand. A reliable power is a generation system that has sufficient power to feed load demand in a period. There exist different methods to express system reliability, where the most popular ones are the loss of load probability (LOLP) and the loss of power supply probability (LPSP)~\cite{Alsadi2018}. In both methods, if the probability is zero, then the load is always fulfilled; otherwise (i.e., probability of one) the load is never fulfilled. LOLP is the probability for the case when a load demand exceeds the generation power by the PV system. On the one hand,  we have a reliable PV system when it can generate sufficient power to fulfill the demanded load within a period. On the other hand, LPSP is defined as the probability of the case when the system generates insufficient power to satisfy the load demand. The main approaches to LPSP demand simulation or probabilistic treatment of time series data to predict dynamic changing on system performance. However, data is not always available and dynamic analysis is complex; and this is a drawback of LOLP and LPSP~\cite{Alsadi2018}.

There exist various methods available related to economic analysis. The main objective is to determine whether the project has an acceptable investment; the usual way is to perform economic analysis after reliability analysis to propose a system with high reliability and lowest cost~\cite{Alsadi2018}. The conventional methods include: Net Present Cost (NPC)~\cite{Park2004}, the Levelized Cost of Energy (LCOE)~\cite{Zhou2010}, or the Life Cycle Cost (LCC)~\cite{Applasamy2011}. The NPC is the present value of all the costs over the project lifetime, minus the present value of all the revenues that it earns over the project lifetime. We find the present net worth by discounting all cash inflows and outflows, including the cost of installation, replacement, and maintenance of the PV system, at an internal rate of return (IRR)~\cite{Park2004}. IRR is used to evaluate the attractiveness of a project or investment. We define LCOE as the average cost per kWh of useful electrical energy produced by the PV system when a lifetime, investment cost, replacement, operation and maintenance, and capital cost are considered~\cite{Kamel2005}. LCOE method is useful in comparing different generation technologies with different operating characteristics~\cite{Zhou2010}. LCC is the estimation of the sum of installation cost, operating and maintenance of a PV system for some time, and expressed in today's value~\cite{Applasamy2011}. Eq.~\eqref{eq:LCC} is used to calculate LCC of a PV system,
\begin{equation}
\label{eq:LCC}
\begin{aligned}
LCC = & C_{PV} + C_{bat} + C_{charger} + C_{inv} + \\
      & C_{installation} + C_{batrep} + C_{PWO\&M},
\end{aligned}
\end{equation}

\noindent where $C_{PV}$ is PV array cost, $C_{bat}$ is initial cost of batteries, $C_{charger}$ is cost of charger, $C_{inv}$ is inverter cost, $C_{installation}$ is installation cost, $C_{batrep}$ is battery replacement cost in present value, and $C_{PWO\&M}$ is operation and maintenance costs 
in present worth.

Parallel to criteria, the designer has to evaluate the design based on optimization variables to recommend an optimal configuration for PV systems. As the number of optimization variables increases, the computational effort increases as well. Hence, to obtain the best PV system design as well as a simplified sizing process, prior work introduced three primary techniques for system sizing calculation, namely intuitive, numerical, and analytical methods~\cite{Zhou2010}. The intuitive method is simple, easy to be implemented, and can be used to give rough suggestions for preliminary design. The sizing rules rely on the designer's experience, using the lowest performance either in a time data or by directly using average value (daily, monthly, or annual) of solar irradiance. This method does not consider the battery's state of charge, or even the random nature of solar irradiation and meteorological conditions~\cite{Alsadi2018}. For the numerical method, the design simulates for each time step within a period to calculate and investigate the state of charge of batteries. The numerical method is very accurate; however, it is complex, demanding more time for calculation~\cite{Park2004}. Analytical methods are used to obtain a close relation or correlation in the form of an equation between capacities and reliabilities. The sizing task becomes much more straightforward than numerical technique; however, the relation cannot be applied to different sites since it is specific to one place of deployment of the PV system, thereby demanding adaptation if another site is analyzed.

\section{Automated Formal Synthesis Method}
\label{sec:Method}

\begin{figure}[h]
\includegraphics[width=0.5\textwidth]{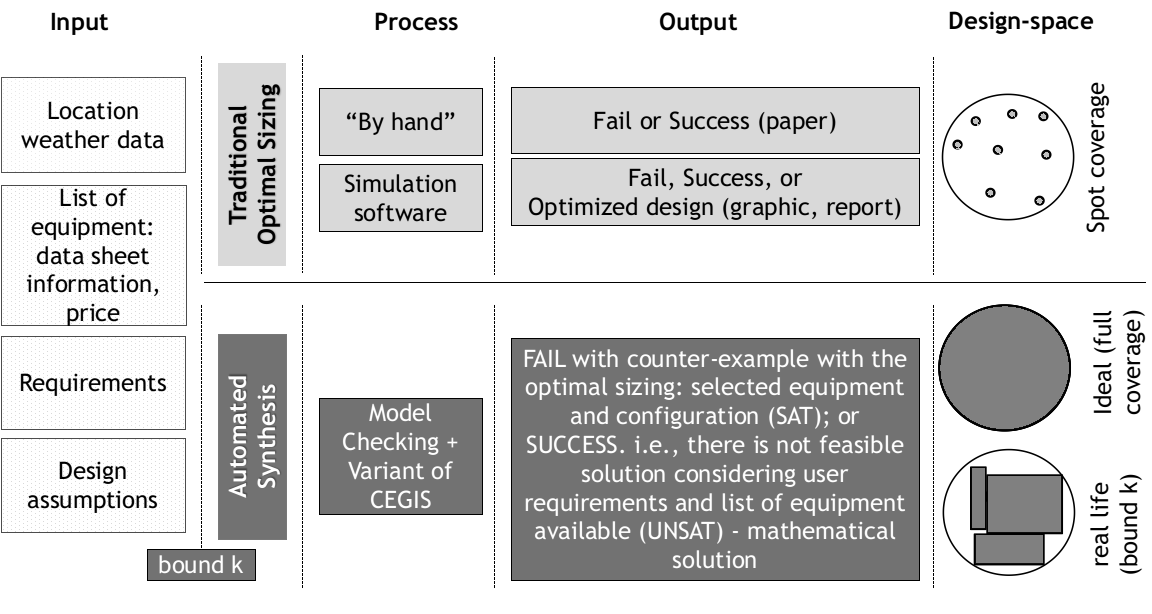}
\centering
\caption{Comparative of optimal sizing methods.}
\label{fig:optimization}
\end{figure}

In this section, we present our theory and methodology to obtain the optimal sizing of stand-alone solar PV systems through program synthesis. In particular, Fig.~\ref{fig:optimization} illustrates how to obtain the optimal sizing of a stand-alone PV system, which shows the traditional techniques (manual and simulation) and the proposed automated synthesis. Note that the input information is the same for all the methods: weather data, price information, design requirements, as load curve and power demand, and design assumptions) except for the automated synthesis where we also define the bound $k$ to restrict the design-space search. Related to the output presented by both techniques, all of them produce a successful or fail result considering a feasible technical solution with the lowest cost. On the one hand, when done by simulation we get a report or graphical result; on the other hand, the automated synthesis technique, which is a mathematical reasoning of a model, presents a counterexample with the optimal solution stored in variables. Furthermore, as described in this section, the design-space coverage during the optimal sizing search is sound and complete when using automated synthesis.

\subsection{Automated Verification Using Model Checking}
\label{sec:AutomatedVerification}

Although simulation and testing explore possible behaviors and scenarios of a given system, they leave open the question of whether unexplored trajectories may contain a flaw. Formal verification conducts an exhaustive exploration of all possible behaviors; when a design is said to be ``correct'' by a formal verification method, it implies all behaviors explored, and questions regarding adequate coverage or missed behavior becomes irrelevant~\cite{Clarke2012}. Formal verification is a systematic approach that applies mathematical reasoning to obtain guarantees about the correctness of a system; one successful method in this domain is model checking~\cite{Clarke2012}. 

To perform the automated formal synthesis, we use a state-of-the-art model checker, awarded with the golden medal at the annual competition in software verification: \textbf{CPAchecker}~\cite{LMU2019}. Automatic program verification requires a choice between precision and efficiency. The more precise a method, the fewer false positives it produces, but also the more expensive it is, and thus applicable to fewer programs. Historically, this trade-off was reflected in two major approaches to static verification: program analysis and model checking. To experiment with the trade-off and to be able to set the dial between the two extreme points, Configurable Program Analysis (CPA) provides a conceptual basis for expressing different verification approaches in the same formal setting. The CPA formalism provides an interface for the definition of program analysis. Consequently, CPAchecker provides an implementation framework that allows the seamless integration of program analysis expressed in the CPA framework. In terms of the architecture, the central data structure is a set of control-flow automata (CFA), which consists of control-flow locations and control-flow edges. The CPA framework provides interfaces to SMT (Satisfiability Modulo Theories) solvers and interpolation procedures~\cite{Beyer2011}. Currently, CPAchecker uses MathSAT as SMT solver~\cite{Beyer2011}.



%
\subsection{Program Synthesis Technique}
\label{sec:ProgramSynthesis}

Program synthesis addresses an age-old problem in computer science: can a computer program itself?~\cite{Bornholt2019}. Before the computer can automatically generate a program, it is necessary to give it a specification of what the program should do. The specification needs to describe the program's desired behavior to ensure that the program does what it intends.

\begin{figure}[h]
	\centering
	\includegraphics[width=0.75\columnwidth]{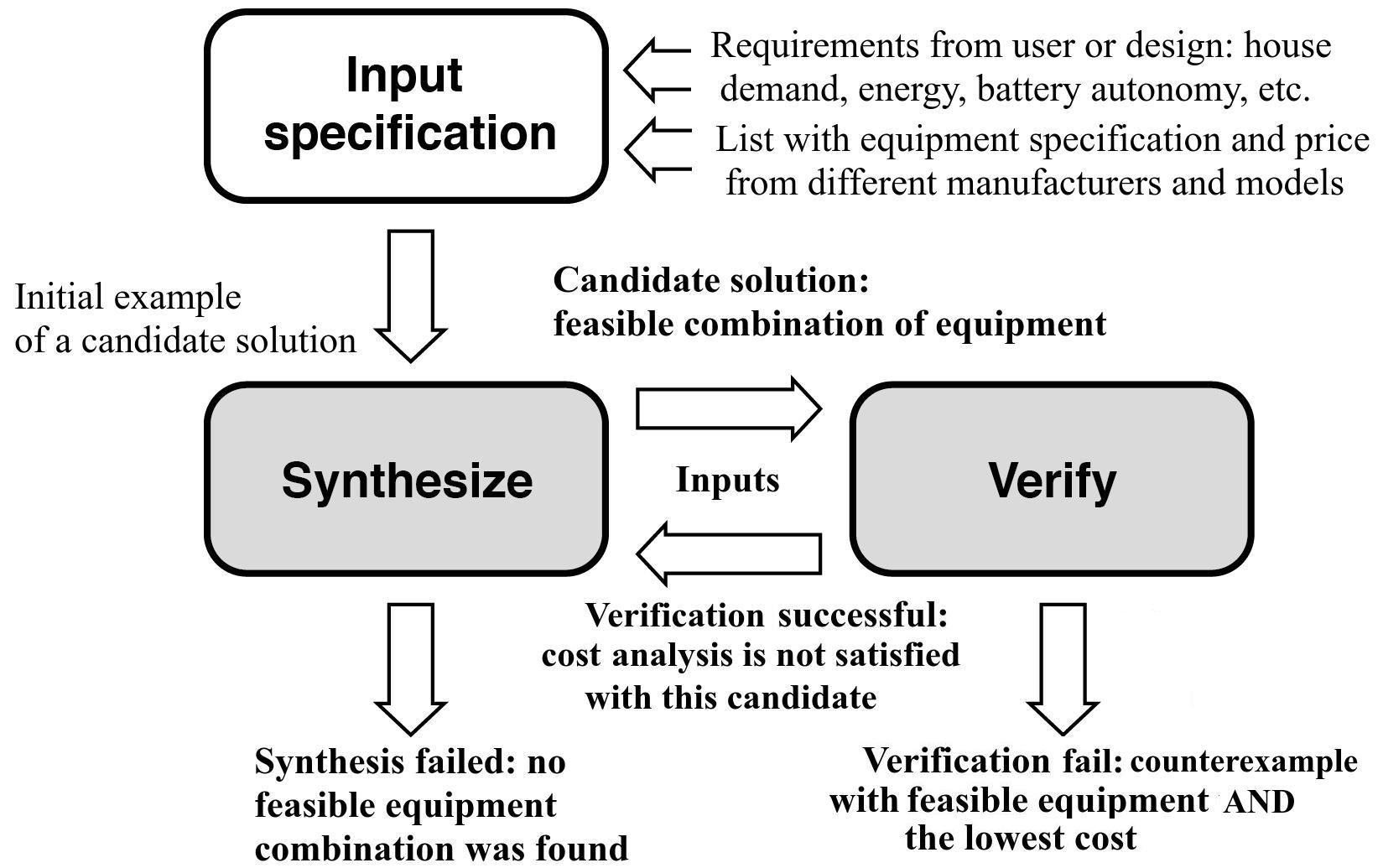}
	\caption{CEGIS adapted to PV system sizing.}
	\label{Counter-Example-Guided-Inductive-Synthesis}
\end{figure}

The basic idea of program synthesis is to automatically construct a program $P$ that satisfies a correctness specification $\sigma$. In particular, program synthesis is automatically performed by engines that use a correctness specification $\sigma$, as starting point, and then incrementally produce a sequence of candidate solutions that partially satisfy $\sigma$~\cite{Abateetal2017}. As a result, a given candidate program $p$ is iteratively refined, to match $\sigma$ more closely. CEGIS represents one of the most popular approaches to program synthesis currently used in practice for cyber-physical problems~\cite{Abateetal2017}, as energy production, distribution, and optimization. Figure~\ref{Counter-Example-Guided-Inductive-Synthesis} illustrates the underlying architecture. Note that CEGIS has close connections to algorithmic debugging using counterexamples and abstraction refinement~\cite{Alur}. 

The correctness specification $\sigma$ provided to our program synthesizer is of the form $\exists \vec{F} . \forall \vec{x}. \sigma(\vec{x}, \vec{F})$, where $\vec{F}$ ranges over functions, $\vec{x}$ ranges over ground terms, and $\sigma$ is a quantifier-free (QF) formula typically supported by SMT solvers. The ground terms are interpreted over some finite domain $\mathcal{D}$, where $\mathcal{D}$ can be encoded using the SMT's bit-vectors part. Examples of specifications used by our method include house demand, energy, and battery autonomy; we also provide a list of equipment specifications and prices from different manufacturers and models.

In Figure~\ref{Counter-Example-Guided-Inductive-Synthesis}, regarding traditional CEGIS method, the phases {\sc Synthesize} and {\sc Verify} interact via a finite set of test vectors {\sc inputs}, which is incrementally updated. Given the correctness specification $\sigma$, the {\sc Synthesize} procedure tries to find an existential witness $\vec{F}$ satisfying the specification $\sigma(\vec{x}, \vec{F})$, for all $\vec{x}$ in {\sc inputs} (as opposed to all $\vec{x} \in \mathcal{D}$). If {\sc Synthesize} succeeds in finding a witness~$\vec{F}$, the latter is a candidate solution (i.e., feasible combination of equipment) to the full synthesis formula, which is passed to {\sc Verify} in order to check whether it is a proper solution ({\it i.e.}, $\vec{F}$ satisfies the specification $\sigma(\vec{x}, \vec{F})$ for all $\vec{x}\in\mathcal{D}$). If this is the case, then the algorithm terminates, i.e., we have found a feasible equipment with the lowest cost; otherwise, in the CEGIS traditional method, additional information is provided to the phase {\sc Synthesize}, in the form of a new counterexample that is added to the {\sc inputs} set and the loop iterates again.

One may notice that each iteration of the traditional CEGIS loop adds a new input to the finite set $\text{\sc inputs}$, which is then used for synthesis. Given that the full set of inputs $\mathcal{D}$ is finite because we use bit-vector expressions, this means that the refinement loop can only iterate over a finite number of times; however, {\sc Synthesize} may conclude that no candidate solution obeying $\sigma$ for the finite set $\text{\sc inputs}$ exists, and our synthesis engine can then conclude that no feasible equipment combination was found.

In our variant CEGIS method, there exist four distinct differences related to the traditional CEGIS: 
(1) there exists no test vector and every candidate is generated during the run-time in the {\sc Synthesize} phase and sent to the {\sc Verify} phase; 
(2) if the {\sc Verify} phase is unsuccessful, then a new candidate is generated by {\sc Synthesize} and 
(3) the lower bound of the {\sc Verify} phase is incremented to search for the lowest cost; 
(4) as a result, there exists no refinement from the {\sc Verify} phase back to the {\sc Synthesize} phase, i.e., 
a new counterexample is not added to the {\sc input} set since a failure during the {\sc Verify} phase will only discard a given candidate, which could be feasible in the next iteration with a new lower bound.

Program synthesis engines that implement the CEGIS approach~\cite{sketch} can automatically produce solutions for a large variety of specifications; here, we have used symbolic software verifiers based on SMT solvers.

Algorithm~\ref{alg:verification-algorithm} describes our pseudo-code to synthesize stand-alone PV systems using symbolic model checking. We adopted the analytical method of optimization, with LCC economical analysis and power reliability based on the critical period criteria.
 \begin{algorithm}
 \caption{Synthesis algorithm}
 \begin{algorithmic}[1]
 \renewcommand{\algorithmicrequire}{\textbf{Input:}}
 \renewcommand{\algorithmicensure}{\textbf{Output:}}
  \STATE Initialize variables \\
  \STATE Declare list of PV panels, controllers, batteries, and inverters data and cost \\
  \STATE Declare the maximum possible cost $MaxCost$  \\
  \STATE Declare power demand, power peak, energy consumption \\
  \STATE Declare battery autonomy, deep of discharge, AC voltage \\
  \FOR {$HintCost=0$ to $MaxCost$}
 	\STATE Non-deterministic variable selects PV Panel from list \\
 	\STATE Non-deterministic variable selects Controller from list \\
 	\STATE Non-deterministic variable selects Battery from list \\
 	\STATE Non-deterministic variable selects Inverter from list \\ 	
 	\STATE Calculate $E_{corrected}, \, E_{p} $ \\
	\STATE Calculate $N_{TPmin}, \, N_{PSmin}, N_{PPmin} $ \\
 	\STATE Calculate $C_{bank}$ \\
	\STATE Calculate $N_{BS}min, \, N_{BP}min, \, N_{B}total$ \\
	\STATE Requirement enforced by \textbf{assume}$(V_{c})$ \\
 	\STATE Calculate $I_{sc,amb}$ \\
 	\STATE Calculate $I_{c,min}$ \\
 	\STATE Requirement enforced by \textbf{assume}$(I_{c} \wedge V_{in}DC \wedge V_{out}AC)$ \\
	\STATE Requirement enforced by \textbf{assume}$(Demand \wedge P_{surge})$ \\
	\STATE Variables hold feasible PV system candidate and cost  \\
	\STATE $F_{obj} \leftarrow  N_{TP}*Panel_{Cost} \, + \, N_{TB}*Battery_{Cost} \, + Controller_{Cost} \, + \, Inverter_{Cost} \, + \, Installation_{Cost} \, + \, batrep_{Cost} \, + \, PWO\&M_{Cost}$ \\
	\STATE Violation check with \textbf{assert}$(F_{obj} > HintCost)$ \\
  \ENDFOR
 \RETURN $(\,)$ 
 \end{algorithmic} 
 \label{alg:verification-algorithm}
 \end{algorithm}
Our synthesis algorithm will synthesize constant values. It starts with the input of manufacturers' data and prices of PV panels, batteries, charge controllers, and inverters (line $2$). After that, we define user requirements, i.e., house requirements and design definitions, from lines $4$ and $5$. 

The \textit{for}-loop on line 6 ensures the search for the lowest cost to the PV solution. The $MaxCost$ value is a very high value put as a limit to the loop, which never will be reached because the optimal solution will be found first. The formal guarantee of this statement is based on the fact that the cost is verified just after a feasible technical candidate if found on {\sc Synthesize} phase. 
In particular, the loop starts with cost $0$ and stops only when the algorithm finds a feasible solution in which the cost breaks the $assertion$ stated in line $22$; if that happens, then our algorithm has found an optimal solution, thereby stating that the {\sc Verify} phase reached a satisfiable condition (\textit{SAT}).

Our synthesis algorithm uses non-deterministic variables to choose one specific constant from a given list of PV panels, controllers, batteries, and inverters (lines $7$ to $10$). That procedure ensures that our synthesis engine checks all combinations of items from each equipment, and combine them to assemble a feasible (candidate) PV solution, which meets the user requirements.

Next, we use Eq.~\eqref{eq:Ecorrected}, Eq.~\eqref{eq:Ep},Eq.~\eqref{eq:NTPmin}, 
Eq.~\eqref{eq:NPSmin}, Eq.~\eqref{eq:NPPmin}, Eq.~\eqref{eq:Cbank}, 
Eq.~\eqref{eq:Nbtotal}, Eq.~\eqref{eq:iscamb}, and Eq.~\eqref{eq:icmin} to calculate the sizing variables (lines $11$ to $17$). The directive \textit{assume} (lines $15$, $18$ and $19$) 
ensures the compatibility of the chosen items from the list of equipment: the {\sc Verify} phase uses only the item (among all the possible ones) that satisfies the statements of Lines $15$, $18$ and $19$. Therefore, our synthesis algorithm reaches line $20$ with one feasible solution, and we calcuate the cost of that solution in $F_{obj}$ (line $21$). 

If our algorithm does not find a feasible solution among the item of equipment that we provide to our {\sc Synthesize} phase, then the result is an unsatisfiable (\textit{UNSAT}), i.e., the program finishes and does not find a solution, which indicates that it was not possible to combine the items of each equipment in order to create a feasible solution. 
The main challenge for the {\sc Synthesize} phase is to find a feasible candidate solution regarding the constraints and user requirements. Related to our {\sc Verify} phase, the challenge is to find the lowest acquisition cost from a list of equipment and components that are provided from the {\sc Synthesize} phase. 
Note that the process described here is completely automated and that validation is performed by our {\sc Verify} phase to ensure that the approach is sound.

\subsection{Assumptions and Premises}

Regarding the line $2$ of Algorithm~\ref{alg:verification-algorithm}, a list of forty equipment from ten different manufacturers was provided to our synthesis engine in order to allow the choice of every item of PV sizing. Data sheet from each item was necessary to collect technical information. Moreover, the price of each item was obtained from available quotations in the market, and if the currency was not in US dollars, then it was used the exchange rate of the day to convert it to US dollars.

With respect to power reliability, this work will rely on the critical period solar 
energy method~\cite{Pinho} as described in Section~\ref{sec:sizing}. 
The usual way is to use loss of load probability (LOLP) or loss of power 
supply probability (LPSP). However, we are neither considering site characteristics nor the load changes over time, 
which demands historical data, the reliability analysis will be developed only 
by the critical period method of PV sizing.
Regarding financial analysis:
\begin{itemize}
	\item LCC lifetime considered: $20$ years;
	\item Installation costs: includes delivery in the isolated community and installation costs itself, $5$\% of total cost~\cite{Agrener2013};
	\item Value of the discount rate or interest rate: $10$\%, which is a good rate considering financial investments in developing countries;
	\item Operation and maintenance annual costs: based on PV projects of similar size in the Amazon region of Brazil, will be adopted the value of US\$ 289.64~\cite{Agrener2013}. This cost includes the battery replacement based on its lifetime ($4$ years for lead-acid batteries), plus inverters and controller replacement (every $10$ years). Therefore, it will be performed three battery bank and one inverter-controller replacements during the LCC analysis.
\end{itemize}

On the subject of PV system optimization technique, we will adopt here the intuitive method 
since the average value daily of solar irradiance is used in the mathematical model, 
without considering the battery's state of charge, or even the random nature 
of solar irradiation and meteorological conditions. Therefore, all the computational 
effort will be concentrated in our automated synthesis algorithm.

Regarding all case studies, it was defined that the minimum state of charge of batteries is $75$\% (with DOD maximum of $25$\%, which is common to lead-acid batteries), the voltage of the system is set in $24$ V DC (the most common as well, but the value can be adjusted to $12$ or $48$ V at the code), and the AC voltage from the inverter is $127$ V (Brazilian standard).

Related to off-the-shelf simulation tools only HOMER Pro perform off-grid system with battery backup analysis and includes economical analysis. Therefore, in this study, HOMER Pro will be the simulation tool used to compare with our automated synthesis method.  Related to HOMER Pro: (a) is available only for Microsoft Windows and its annual standard subscription costs US\$ $504.00$~\cite{HOMER}; (b) it does not have the LCC cost in its reports. It has NPC and LCOE. Therefore NPC was used to obtain LCC in order to allow the comparative; (c) the optimization analysis of HOMER allows to define a load curve and temperature according of data collected from online databases. However, in order to allow a correct comparative, the curve load and the temperature were defined exactly the same as automated synthesis tools; (d) it does not have a explicit equipment called charge controller. It uses a controller resource that can perform in two different ways, according of the optimization choice or the user choice: load following or cycle charging~\cite{HOMER}. During the tests it was chosen the load following controller: it produces only enough power to meet the demand~\cite{HOMER}; (e) It was assumed the value of 5\% of capacity shortage that is equivalent to 95\% of availability of the PV system. By definition, availability is the percentage of time at which a power system is capable of meeting the load requirements~\cite{Khatib2014}. For critical loads, 99\% is considered acceptable. While in a ordinary house electrical load, 95\% is considered acceptable; (f) it was assumed a string of two batteries in order to match the voltage of the system of $24$ V DC that was used for the automated synthesis tool; (g) the premise adopted when using HOMER Pro it was that the user does not know the optimal solution, and that in order to obtain this solution is necessary to include (at the design phase of the tool) generic PV and batteries modules that HOMER will search for the optimized power of each component. With that in mind, it was included a generic flat plate PV of $1$ kW and generic lead-acid batteries of $1$ kW as well (and with capacity of $83.4$ Ah according with HOMER Pro modeling). HOMER, during run-time, decides the size in kW of each module, based on feasibility and lower cost.

\section{Case Studies and Analysis}
\label{sec:Results}

This section describes our case studies, experimental setup,  objectives, and results to evaluate our proposed synthesis approach. We also compare our approach with a specialized simulation tool (HOME Pro).

\subsection{Case studies} 

We have performed seven case studies to evaluate our proposed synthesis approach, as described in the first column of Table~\ref{tab1} (\textit{Specification}). These case studies were defined based on real houses visited by the team of a Newton Fund project in riverside communities around the Low Black River in Amazonas - Brazil. This project finished in March 2019, where we visited and surveyed 14 riverside isolated communities, aiming to evaluate the energetic habits of the dwellers.\footnote{http://star-energy.coventry.ac.uk/} For all cases, an estimated load curve (kWh) was defined based on the electronics consumers found of each house.

\begin{table}[!t]
\caption{Case studies and results: optimization of PV systems.}\label{tab1}
\begin{scriptsize}
\begin{tabular}{|c|c|c|}
\hline
\hline
Tools & \makecell{CPAchecker 1.8\\(MathSAT 5.5.3)}& HOMER Pro 3.13.1\\
\hline
\hline
Specification & Result & Result \\
\hline
\makecell{\textbf{Case Study 1}\\Peak:342W\\Surge:342W \\E:3,900Wh/day\\Autonomy:48h} & \makecell{SAT (172.03 min) \\NTP:1$\times$340W (1S)\\NBT:8$\times$105Ah (2S-4P)\\Controller 15A/75V\\Inverter 700W/48V\\LCC: US\$ 7,790.53} & \makecell{(Time: 0.33 min)\\2.53 kW of PV\\NBT:12$\times$83.4Ah (2S-6P)\\0.351kW inverter\\LCC: US\$ 7,808.04}\\
\hline
\makecell{\textbf{Case Study 2}\\Peak:814W\\Surge:980W\\E:4,880Wh/day\\Autonomy:48h} & \makecell {SAT (228.7 min) \\NTP:2$\times$330W (2S)\\NBT:10$\times$105Ah (2S-5P)\\Controller 20A/100V DC\\Inverter 1,200W/24V \\LCC: US\$ 8,335.90} & \makecell{(Time: 0.18 min)\\3.71 kW of PV\\NBT:20$\times$83.4Ah (2S-10P)\\0.817kW inverter\\LCC: US\$ 12,861.75} \\
\hline
\makecell{\textbf{Case Study 3}\\Peak:815W\\Surge:980W\\E:4,880Wh/day\\Autonomy:12h} & \makecell {SAT (166.13 min) \\NTP:4$\times$150W (4S)\\NBT:4$\times$80Ah (2S-2P)\\Controller 15A/100V DC\\Inverter 1,200W/24V \\LCC: US\$ 7,306.27} & Not possible \\
\hline
\makecell{\textbf{Case Study 4}\\Peak:253W\\Surge:722W\\E:3,600Wh/day\\Autonomy:48h} &  \makecell {SAT (143.71 min) \\NTP:4$\times$150W (4S)\\NBT:10$\times$80Ah (2S-5P)\\Controller 15A/75V\\Inverter 750W/24V \\LCC: US\$ 7,816.31} & \makecell{(Time: 0.23 min)\\2.42 kW of PV\\NBT:12$\times$83.4Ah (2S-6P)\\0.254kW inverter\\LCC: US\$ 7,677.95}\\
\hline
\makecell{\textbf{Case Study 5}\\Peak:263W\\Surge:732W\\E:2,500Wh/day\\Autonomy:48h} &  \makecell {SAT (134.93 min) \\NTP:1$\times$340W (1S)\\NBT:6$\times$105Ah (2S-3P)\\Controller 15A/75V\\Inverter 400W/24V \\LCC: US\$ 7,252.14} & \makecell{(Time: 0.18 min)\\1.59 kW of PV\\NBT:10$\times$83.4Ah (2S-5P)\\0.268kW inverter\\LCC: US\$ 6,175.57} \\
\hline
\makecell{\textbf{Case Study 6}\\Peak:322W\\Surge:896W\\E:4,300Wh/day\\Autonomy:48h} &  \makecell {SAT (235.75 min) \\NTP:2$\times$200W (2S)\\NBT:10$\times$105Ah (2S-5P)\\Controller 15A/75V\\Inverter 400W/24V \\LCC: US\$ 8,287.23} & \makecell{(Time: 0.22 min)\\3.15 kW of PV\\NBT:14$\times$83.4Ah (2S-7P)\\0.328kW inverter\\LCC: US\$ 9,112.45} \\
\hline
\makecell{\textbf{Case Study 7}\\Peak:1,586W\\Surge:2,900W\\E:14,000Wh/day\\Autonomy:48h} & TO & \makecell{(Time: 0.20 min)\\12.5 kW of PV\\NBT:66$\times$83.4Ah (2S-33P)\\1.60kW inverter\\LCC: US\$ 41,878.11} \\
\hline
\hline
\end{tabular}
\\Caption: OM = out of memory; TO = timeout; IF = internal failure, E = energy.
\end{scriptsize}
\end{table}

\subsection{Setup} 

The start-of-art verification tool CPAchecker\footnote{Command-line: \$ scripts/cpa.sh -heap 64000m -config config/bmc-incremental.properties -spec config/specification/sv-comp-reachability.spc filename.c} was used as our verification engine to compare our approach effectiveness and efficiency. The Simulation tool HOMER Pro version $3.13.1$ was used for comparative purpose.

All experiments regarding the verification tools were conducted 
on an otherwise idle Intel Xeon CPU E5-4617 ($8$-cores) with 
$2.90$ GHz and $64$ GB of RAM, running Ubuntu $16.04$ LTS $64$-bits. 
For HOMER Pro, we have used an Intel Core i5-$4210$ ($4$-cores), 
with $1.7$ GHz and $4$ GB of RAM, running Windows 10. 
Our experiments were performed with a predefined timeout of $240$ minutes.

\subsection{Objectives} 

Our evaluation aims to answer two experimental questions: 

\begin{enumerate}

\item[EQ1] \textbf{(soundness)} does our automated synthesis approach provide correct results?

\item[EQ2] \textbf{(performance)} how does our formal synthesis tool compare to a specialized simulation tool?

\end{enumerate}

\subsection{Results}  

CPAchecker was able to synthesize the optimal sizing in six out of seven case studies: the result was produced within the time limit, which varied from $134.71$ to $235.75$ minutes. Only case study $7$ led to a \textit{timeout} result, i.e., it was not solved within $240$ minutes. However, if we remove this timeout limitation from CPAchecker, the verifier can solve the optimization in $44.97$ hours. The violation (SAT result) indicated in Table~\ref{tab1} is the $assert$ of line $22$ from Algorithm~\ref{alg:verification-algorithm}. As an interesting feature of the automated synthesis, we present every optimal sizing as a detailed list of equipment with brand (in Table~\ref{tab1} it was removed to avoid some advertising) and model taken from the list provided to the algorithm. Moreover, we present the number of series and parallel solar panels and batteries.

Related to HOMER Pro, it was able to evaluate six case studies, and within a time shorter than $30$ seconds, which was much faster than our automated synthesis tool (cf.~\textit{EQ2}). We were unable to simulate case study $3$ since HOMER Pro does not have the feature of adjusting the battery autonomy (the tool always meets the user requirement, i.e., the load curve during the $365$ days of the year). Other HOMER Pro drawbacks, when compared to our automated synthesis method: (a) There exists no explicit charge controller as a system equipment. HOMER includes a controller automatically just to simulate the charge/discharge of batteries and to meet the load requirement; however, without costs or even with electrical characteristics as maximum current and voltage, which are common during PV sizing; (b) HOMER demands to include some battery specification to initiate the optimization; however, it does not change the electrical specifications during the simulation; the presented results are multiples of the original battery type suggested by the user. As example, it was started with a $83.4$ Ah lead-acid battery, and during the simulation, HOMER Pro did not try to use other capacities or types; (c) HOMER does not present the optimal solution in terms of connections of arrays of PV panels, just the total in terms of power, i.e., it does present neither models and the power of each PV panel nor the total of panels in series or parallel; (d) Battery autonomy is not a parameter that the user can set when using HOMER. 

Comparing the results between the formal synthesis with CPAchecker and HOMER Pro (cf.~\textit{EQ2}), 
we observed that most results are quite similar, in terms of technical solution and cost (cf. Table~\ref{tab1}). 
Mainly related to LCC, the cost was very close in cases $1$, $4$, $5$ and $6$, with difference varying from $0.23$\% to $17.4$\%. 
Even adopting the same price per kW to the PV panels, 
inverters, and batteries, HOMER Pro does not use costs 
related to charge controllers, which we introduced into the 
CPAchecker modeling. The premise used in CPAchecker to adopt 
a fixed annual cost for operation and maintenance can produce 
some impact as well at this discrepancy; however, it is not significant
since this annual cost is too small when compared to the resulting LCC value.
However, there exists a considerable divergence in case study $2$, where the costs presented by HOMER Pro were $54$\% higher than our automated synthesis tool, probably because we underestimated the operation and maintenance costs assumed by our automated synthesis tool to that specific load. 

In general, the size of the PV panels and battery bank were larger in HOMER Pro than with our formal synthesis approach, and determining which one is the correct depends on a comparison with field-deployed solar systems. The mathematical models are different in our model and the simulation tool, and particular parameters can be tuned. This may explain the difference presented in all case studies. As comparative, consider case study $1$: the optimal solution provided by HOMER Pro demands $7$ $\times$ more PV panels than the solution presented by our synthesis tool, and HOME Pro does not show the arrangement of arrays (i.e., the number of series and parallel PV panels); the battery bank presented by HOMER Pro provides $500.4$ Ah of capacity ($6 \times 83.4$), while our synthesis tool presented an optimal solution with $420$ Ah of total capacity ($4 \times 105$). 

To compare the results obtained from the optimization with the real-world, the authors had four PV systems deployed and monitored since June $2018$ in a riverside community in the Amazonas State in Brazil (coordinates 2$^{o}$44'50.0"S 60$^{o}$25'47.8"W), with similar power demands presented by case studies $1$, $4$, $5$, and $6$, 
always with a $3$ $\times$ $325$ W ($3$S) panels and $4$ $\times$ $220$ Ah ($2$S-$2$P $= 440$ Ah) lead-acid batteries. These solutions are closer to the result presented by our formal synthesis approach than HOMER Pro, thereby showing that our solution is sound, which answers \textit{EQ1}.

Related to the inverters, HOMER Pro suggests a value in kW very close to the peak of every case study, and it is just a reference value and not a commercial value of the employed inverter. Our synthesis tool, however, 
presents inverters that are commercial and can be found off-the-shelf; therefore, our approach is a PRO to the formal synthesis method.

Concerning the charge controllers, as we reported in the previous section, HOMER Pro does not include it as an explicit equipment in its mathematical model; only our synthesis tool presents a commercial controller and includes it during the cost analysis. Therefore, the formal synthesis method presents more reliable results than HOME Pro.

Our synthesis tool did not solve the case study $7$  within the time limit established during the experimental phase. Case study $3$ was not possible to simulate in HOMER Pro, because its restriction does not allow one to set the battery autonomy, thus resting both without parameters for comparison.

\subsection{Threats to validity}

We have reported a favorable assessment of our formal synthesis method to obtain the optimal size of the PV system. 
Nevertheless, we have also identified three main threats to the validity of our experimental results, which can be further assessed and 
constitute future work: ($1$) improvement of the power reliability 
analysis: to include loss of load probability or loss of power 
supply probability, which can make the analysis more precise by considering the dynamic of the weather characteristics over the year or by electric load changes over time, 
based on historical data; 
($2$) the cost analysis is well-tailored to the Amazon region of Brazil; 
however, a broad analysis from other isolated areas must be 
performed to make the optimization general in terms 
of applicability in other isolated areas of the world, such as India and China; ($3$) to deploy at the field some PV systems 
sized using our synthesized results to validate it since a comparative with a real system may be more reliable than comparing with a simulation tool.

\section{Conclusion and Future Work}
\label{sec:Conclusion}

We have described and evaluated an automated synthesis method to obtain the optimal size of the PV system using software model checking techniques. The focus was on the synthesis method to obtain the optimal solution based on formal methods, which can cover better the design-space as opposed to simulation tools. We have considered seven case studies from PV systems in two different sites of the Amazonas State in Brazil, ranging from $253$\,W to $1,586$\,W peak; one state-of-art verification engine was considered (CPAchecker), in addition to a specialized off-the-shelf simulation tool (HOMER Pro) to compare the results.

The paper produced a methodological research with innovative value regarding the first use of automated synthesis for optimal sizing of solar PV systems.
In summary, our synthesis tool is capable of presenting a solution, which is far detailed and close to the commercial reality than the solution presented by HOMER Pro. In particular, our method can provide all the details of every component of a PV system solution, with complete electrical details from data sheet of manufacturers, including the model of the component, nominal current, and voltage; we also cover the charge controller, which is unavailable in HOMER Pro. Note that our automated synthesis tool took longer to find the optimal solution than HOMER Pro; however, the presented solution is sound and complete; it also provides a list of equipment to be bought from manufacturers. Moreover, we extended the CEGIS synthesis method and implemented this extension within our proposed formal synthesis tool, which allows the optimization of stand-alone PV systems with the best compromise between power reliability and system cost analysis.

For future work, we plan to improve the power reliability analysis, to address the restriction to only allow automated synthesis of riverside communities in the Amazonas state (Brazil) and to validate some cases with the deployment of real PV systems in isolated communities.

\section*{Acknowledgment}

The authors would like to thank to University of Sheffield's QR GCRF for HOMER Pro license.

\ifCLASSOPTIONcaptionsoff
  \newpage
\fi

%
\bibliographystyle{IEEEtran}
\bibliography{trindadeThesis}{}
\end{document}